\documentclass[pre,twocolumn,amsmath,amssymb]{revtex4}

\usepackage{graphicx}
\usepackage{amssymb}
\usepackage{amsmath}
\usepackage{enumerate}

\usepackage{url}
\usepackage[colorlinks,urlcolor=magenta]{hyperref}

\begin{document}

\title{A Reaction-Diffusion Model of the Cadherin-Catenin System: A Possible Mechanism for Contact Inhibition and Implications for Tumorigenesis}

\author{Markus Basan}
\affiliation{Laboratoire Physico-Chimie Curie, CNRS-UMR 168, Universit\'e Pierre et Marie Curie Paris $\rm V\!I$,
Institut Curie Centre de Recherche, Paris, France}

\author{Timon Idema}
\affiliation{Instituut-Lorentz for Theoretical Physics, Leiden Institute of Physics, Leiden University, Leiden, The Netherlands}

\author{Martin Lenz}
\affiliation{Laboratoire Physico-Chimie Curie, CNRS-UMR 168, Universit\'e Pierre et Marie Curie Paris $\rm V\!I$,
Institut Curie Centre de Recherche, Paris, France}

\author{Jean-Fran\c{c}ois Joanny}
\affiliation{Laboratoire Physico-Chimie Curie, CNRS-UMR 168, Universit\'e Pierre et Marie Curie Paris $\rm V\!I$,
Institut Curie Centre de Recherche, Paris, France}

\author{Thomas Risler}
\affiliation{Laboratoire Physico-Chimie Curie, CNRS-UMR 168, Universit\'e Pierre et Marie Curie Paris $\rm V\!I$,
Institut Curie Centre de Recherche, Paris, France}

\begin{abstract}
Contact inhibition is the process by which cells switch from a motile growing state to a passive and stabilized state upon touching their neighbors. When two cells touch, an adhesion link is created between them by means of transmembrane E-cadherin proteins. Simultaneously, their actin filaments stop polymerizing in the direction perpendicular to the membrane and reorganize to create an apical belt that colocalizes with the adhesion links. Here, we propose a detailed quantitative model of the role of the cytoplasmic $\beta$-catenin and $\alpha$-catenin proteins in this process, treated as a reaction-diffusion system. Upon cell-cell contact, the concentration in $\alpha$-catenin dimers increases, inhibiting actin branching and thereby reducing cellular motility and expansion pressure. This model provides a mechanism for contact inhibition that could explain previously unrelated experimental findings on the role played by E-cadherin, $\beta$-catenin and $\alpha$-catenin in the cellular phenotype and in tumorigenesis. In particular, we address the effect of a knockout of the adenomatous polyposis coli tumor suppressor gene. Potential direct tests of our model are discussed.
\end{abstract}

\maketitle

\section*{Introduction}

\begin{figure}[h]
\includegraphics[width=\linewidth]{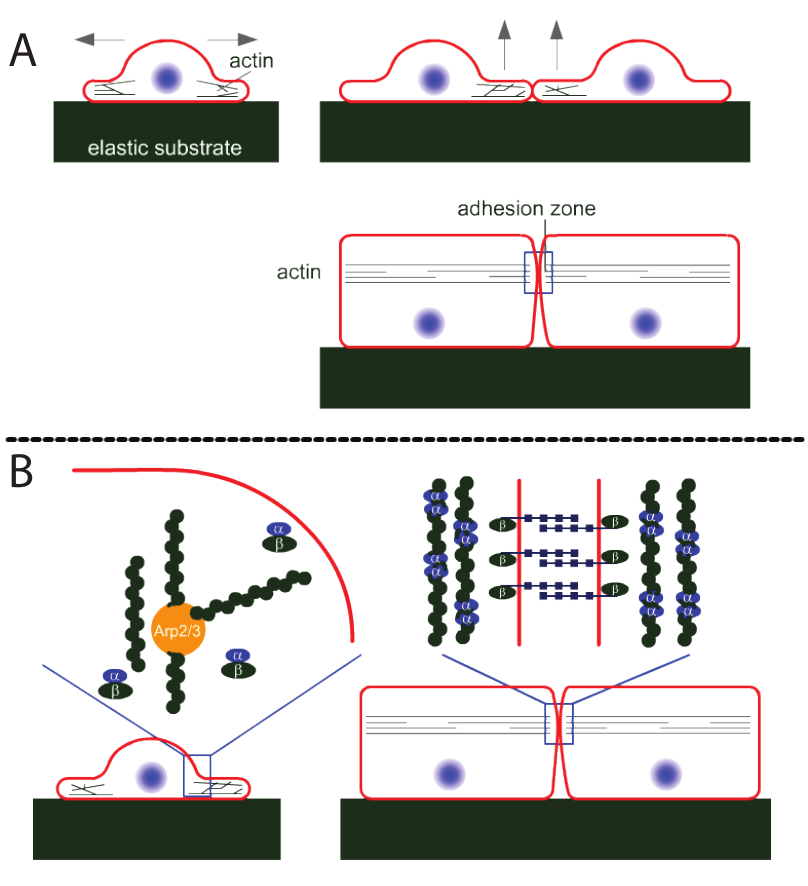}
\caption {\label{fig_adhesion}
(A) Schematic illustration of the establishment of the epithelial cell-cell adhesion zone. After cells have spread via protrusions along the substrate and become confluent, they start growing upwards and colocalize their actin belt while forming the adhesion zone.
(B) Different organizations of cortical actin, $\alpha$-catenin and $\beta$-catenin-related complexes in epithelial cells during their spreading ({\it left}) and after the mature epithelial sheet has been formed ({\it right}). Before cell-cell contact, $\beta$-catenin-E-cadherin complexes are present in the cytoplasm and therefore recruit $\alpha$-catenin proteins before they can form dimers, which lets Arp2/3 complexes branch the actin network. In the presence of a neighboring cell however, $\beta$-catenin-E-cadherin complexes are mostly found at the cell membrane, which favors the formation of $\alpha$-catenin dimers in the cytoplasm. These dimers further bind strongly to actin, effectively excluding Arp2/3 complexes from the actin network and favoring parallel bundling.
}
\end{figure}

Before the establishment of cell-cell contacts, epithelial cells are in a motile and growing state. The polymerizing actin filaments create forces on the membrane that are responsible for the formation of lamellipodia and filopodia~\citep{peskin1993cma, vasioukhin2000dap}. Moreover, the actin filaments undergo continuous branching and growth resulting in dynamic extensions of the membrane~\citep{pollard2003cellular}. When cells are scarce and do not contact each other, E-cadherins are found both on the plasma membrane and in membrane vesicles within the cytoplasm, but their role is minimal: when located on the membrane, they quickly get endocytosed into cytoplasmic vesicles~\citep{bryant04}. After they have grown enough to cover the substrate in a confluent layer, epithelial cells become polarized perpendicular to the substrate. At this point, they no longer produce lamellipodia and filopodia, but instead reorganize their actin into a belt located near their apical side (see Fig. \ref{fig_adhesion}~A)~\citep{hirokawa1983mechanism}.
Simultaneously, the E-cadherins located in the plasma membrane link their extracellular domains with the cadherins of the neighboring cells and colocalize with the actin belt, forming what is known as the adhesion zone. The linkage of E-cadherins stabilizes their localization on the plasma membrane, effectively depleting them from the cytoplasm~\citep{bryant04, de2009endocytosis}. 

The reorientation of the actin filaments upon cell-cell contact indicates a reduced activity of branching proteins such as Actin-related proteins 2 and 3 (Arp2/3) and an increased activity of bundling proteins such as $\alpha$-catenin dimers (see Fig.~\ref{fig_adhesion}~B). When oriented parallel to it, the growing actin filaments no longer exert a force on the plasma membrane. Therefore, the cell downregulates both its motility and expansion pressure in response to reaching confluence, a process referred to as contact inhibition.

In 2005, Drees et al.~\citep{drees2005alphacm} challenged the textbook view that $\alpha$-catenin mechanically links the adhesion complex to the underlying actin cytoskeleton. They showed that $\alpha$-catenin exists either as a monomer or as a dimer, and that the domain on an $\alpha$-catenin monomer that binds to $\beta$-catenin and the one that binds to another $\alpha$-catenin monomer overlap. Therefore, the formations of $\alpha$-catenin dimers and $\alpha$-catenin-$\beta$-catenin complexes are mutually exclusive~\citep{yamada2005dcc}. Dimeric $\alpha$-catenin can bundle actin filaments and competes for actin binding sites with Arp2/3. According to these findings, a high concentration of $\alpha$-catenin dimers therefore suppresses actin branching, growth and expansion pressure (see reviews~\citep{gates2005can, lecuit2007cell} and Fig.~\ref{fig_adhesion}~B).

A loss of contact-inhibition via epithelial-mesenchymal transition is an essential step for tumorigenesis~\citep{thiery2002epithelial}. It has recently been proposed that an excess expansion pressure could be a characteristic trait of neoplastic tissues~\cite{basan09}. A breakdown of the regulation mechanism discussed above might therefore lead to tumorigenesis. It is indeed well-known that the E-cadherin-$\beta$-catenin-$\alpha$-catenin adhesion complex plays an important role in carcinomas~\citep{hirohashi2003cas, conaccisorrell2002cca}. A reduced expression of E-cadherins---for example due to DNA hypermethylation---is associated with a loss of cellular polarity and the acquisition of invasive characteristics~\citep{hirohashi1998icm}. However, it has been shown that overexpression and reduced degradation of $\beta$-catenins also leads to cellular transformations that result in the cell's ability to grow in soft agarose gels and to overproliferate at high cell densities \citep{orford1999eebeta}. Along the same lines, in cells that have undergone the epithelial-mesenchymal transition, E-cadherin expression is downregulated while the production of $\beta$-catenin is increased~\citep{eger2000emt}. It has also been shown that the growth-inhibiting activity of E-cadherin is counteracted by an increased $\beta$-catenin activity~\citep{stockinger2001crc}. Finally, the concentration of $\beta$-catenins is regulated by the adenomatous polyposis coli (APC) protein, a tumor suppressor protein that is known to label $\beta$-catenin for degradation~\citep{korinek1997cta}. On the other hand, $\beta$-catenin-null cells show an unaffected or even decreased rate of expansion and proliferation~\citep{posthaus2002betacn, fevr2007wbetac}. As the important role of E-cadherin and $\beta$-catenin in the progression of cancer has been well studied, several papers report that the loss of $\alpha$-catenin is an important prognostic factor for cancer, as reviewed in Benjamin and Nelson~\citep{benjamin2008bench}. For example, the ablation of $\alpha$-catenin in the skin causes cellular hyperproliferation, occurrence of mitoses away from the basal layer and defects in epithelial polarity \citep{vasioukhin2001had}. These phenotypes are remarkably similar to those obtained with a modified expression level of E-cadherin or $\beta$-catenin proteins.

Although strong indications exist that the influence of $\alpha$-catenin on actin polymerization plays a role in contact inhibition, the functional details of this mechanism remain unclear. Important progress has been made in this direction by Drees and coworkers~\citep{drees2005alphacm, yamada2005dcc}, who proposed a picture in which cell-cell contact leads to an accumulation of E-cadherin-$\beta$-catenin-$\alpha$-catenin complexes at the adhesion sites. They propose that the release of $\alpha$-catenin monomers from these complexes into the cytoplasm provide an increase in $\alpha$-catenin dimer concentration, favoring actin bundling and downregulating actin assembly and branching. In the present work, we propose a model for the E-cadherin-$\beta$-catenin-$\alpha$-catenin function that is based on a reaction-diffusion system. We show that the interplay between these three proteins results in a pathway for contact inhibition that downregulates actin polymerization in response to cell-cell contact.

Our mechanism relies on the fact that the binding of $\beta$-catenin to $\alpha$-catenin limits $\alpha$-catenin dimerization in the cytosol. When $\beta$-catenin-E-cadherin complexes are recruited to the cell membrane due to cell-cell contact, the cytosolic concentration of $\beta$-catenin drops and $\alpha$-catenin dimerization can take place. According to the work by the Nelson and Weis group~\citep{drees2005alphacm}, this in turn prevents Arp2/3-based actin branching and cause the cell to enter a quiescent state. Using the framework of our physical model, we investigate the effect of disruptions of this pathway and obtain results that are consistent with experimentally observed cellular transformations that lead to tumorigenesis.

\section*{Results}

\subsection*{Description}

\begin{figure}[h]
\includegraphics[width=0.7\linewidth]{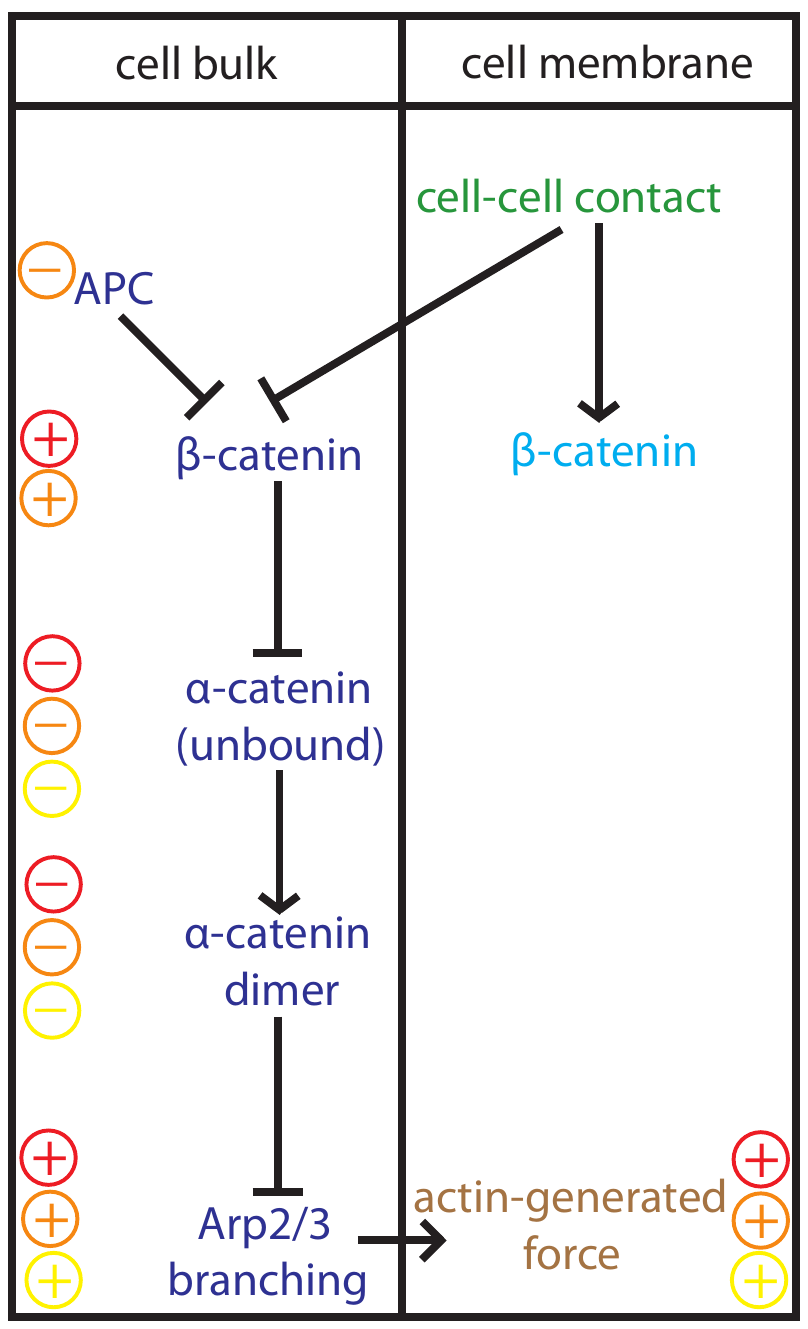}
\caption {\label{fig_pathway}
Schematic illustration of the proposed cadherin-catenin pathway for contact inhibition as well as its possible disruptions. Arrows and T-bars between the different gene, proteins or cell states of the diagram indicate induction and repression, respectively. The signs of different colors illustrate how various events can lead to a breakdown of this pathway: reduced expression of E-cadherins ({\it red, upper signs}), mutation of the APC tumor suppressor gene ({\it orange, middle signs}), and reduced expression or mutations of $\alpha$-catenin ({\it yellow, lower signs}). Minus signs indicate either decreased concentrations or complete impairment of the associated proteins, and plus signs indicate increased concentrations as compared with the healthy cell state.
}
\end{figure}

The mechanism of the pathway we propose is schematically illustrated in Fig.~\ref{fig_pathway}.
Its main feature is that $\alpha$-catenin-$\beta$-catenin binding competes with $\alpha$-catenin dimerization: At high cytosolic concentrations of $\beta$-catenins, the majority of $\alpha$-catenins enters $\alpha$-catenin-$\beta$-catenin complexes. At low cytosolic concentrations of $\beta$-catenins however, $\alpha$-catenins form dimers almost exclusively~\citep{drees2005alphacm}. Therefore, the organization and activity of the actin cortex of the cell depends on the presence of a neighboring cell by the following mechanism: It is known that $\beta$-catenin quickly binds to E-cadherin after production at the Golgi apparatus of the cell~\citep{hinck94}. When the cell is in its growth phase, E-cadherin-$\beta$-catenin complexes are mostly found in vesicles in the cytoplasm~\citep{bryant04}, effectively creating a large concentration of $\beta$-catenin complexes in the cytosol. These complexes further recruit most of the $\alpha$-catenin monomers that are present in the cytoplasm, letting actin binding sites free for Arp2/3 complexes to bind. The structure of the actin cortex is therefore branched, and its activity is high. In contrast, when contact with a neighboring cell is established, the E-cadherins bind to the neighboring cell and accumulate at the membrane, effectively lowering their concentration in the cytosol~\citep{le1999rcp}. Since a large fraction of E-cadherins is bound to $\beta$-catenins, the establishment of cell-cell contacts also induces a redistribution of $\beta$-catenins to the plasma membrane. Indeed, the potent ability of E-cadherin to recruit $\beta$-catenin to the cell membrane has been observed in vivo~\citep{orsulic1999cbp}. This $\beta$-catenin redistribution to the plasma membrane in turn favors the formation of $\alpha$-catenin dimers in the cytosol, which further favors actin bundling rather than actin branching and polymerization. Note that other protein complexes could play a similar role as E-cadherins and transport $\beta$-catenin proteins to the cell membrane upon cell-cell contact, as it has been proposed recently~\cite{maher2009activity}. In any case, the cell switches from an active state with high actin branching and polymerization activity when there is no cell-cell contact, to a passive state characterized by a reduced actin activity after cell-cell contact has been established.

\subsection*{Model Equations}

To model the cadherin-catenin system described above in a quantitative manner, we write a system of reaction-diffusion equations for the cytosolic concentrations of the different proteins involved. (Note that active transport of these proteins may be involved but that we do not expect the mechanism presented in this paper to depend crucially on this aspect.) In this model, we treat all protein bindings as irreversible because binding affinities are high (typically with energies of many $k_{\rm B} T$~\citep{pokutta2000sda}). Assuming that the protein production rates and the configuration of neighboring cells are constant in time, we can focus on the steady-state dynamics of the system, which for the cytosol of the cell can be written as
\begin{subequations} \label{bulkdiffusion}
\begin{eqnarray}
D_{\alpha} \nabla^2 C_{\alpha} - k_{\alpha \beta} C_{\alpha} C_{\beta} - 2\,k_{\alpha \alpha} C_{\alpha}^2 - r_{\alpha} C_{\alpha} &=& 0 \label{Eq1}\\
D_{\beta} \nabla^2 C_{\beta} - k_{\alpha \beta} C_{\alpha} C_{\beta} - r_{\beta} C_{\beta} &=& 0 \label{Eq2}\\
D_{\alpha \alpha} \nabla^2 C_{\alpha \alpha} + k_{\alpha \alpha} C_{\alpha}^2 - \left( r_{\alpha \alpha} + \tilde{r}_{\alpha \alpha} \right) C_{\alpha \alpha} &=& 0 \label{Eq3}\\
D_{\alpha \beta} \nabla^2 C_{\alpha \beta} +  k_{\alpha \beta} C_{\alpha} C_{\beta} - r_{\alpha \beta} C_{\alpha \beta} &=& 0. \label{Eq4}
\end{eqnarray}
\end{subequations}
These equations respectively describe the diffusion dynamics of $\alpha$-catenin, $\beta$-catenin, $\alpha$-catenin dimers and $\alpha$-catenin-$\beta$-catenin complexes in the cytoplasm.
Here  $C_{\alpha}$, $C_{\beta}$, $C_{\alpha \alpha}$ and $C_{\alpha \beta}$ are the cytoplasmic protein concentrations of $\alpha$-catenin monomers, $\beta$-catenin monomers (bound to cytosolic E-cadherins), $\alpha$-catenin dimers, and $\alpha$-catenin-$\beta$-catenin complexes (bound to cytosolic E-cadherins), respectively; $D_{\alpha}$, $D_{\beta}$, $D_{\alpha \alpha}$ and $D_{\alpha \beta}$ are the associated diffusion constants and $r_{\alpha}$, $r_{\beta}$, $r_{\alpha\alpha}$ and $r_{\alpha\beta}$ the associated degradation rates; $k_{\alpha\alpha}$ and $k_{\alpha\beta}$ are respectively the rates of $\alpha$-catenin dimerization and $\alpha$-catenin-to-$\beta$-catenin binding; and $\tilde{r}_{\alpha \alpha}$ is the reaction rate of $\alpha$-catenin dimers with actin. Note that since most $\beta$-catenins bind to E-cadherins immediately after production~\citep{hinck94}, we do not explicitly model the reaction-diffusion dynamics of E-cadherins but instead account for its important effect on the redistribution of $\beta$-catenin-E-cadherin complexes in the boundary conditions at the plasma membrane, as we shall see below. Modeling the diffusion and reactions of E-cadherins and $\beta$-catenins separately adds another layer of complexity to our model, but would not qualitatively change our main results. Therefore, when we refer to $\beta$-catenin in our model, we implicitly mean the E-cadherin-$\beta$-catenin complex. Note also that the effect of the Wnt signaling pathway on $\beta$-catenin is taken into account effectively in the bulk degradation rate of this protein.

Production of these proteins in the vicinity of the cell nucleus as well as their interactions with the plasma membrane need to be accounted for using appropriate boundary conditions.
The production of $\alpha$-catenin and $\beta$-catenin in the Golgi apparatus of the cell is taken into account by fixed influxes of proteins into the cytoplasm, denoted by $j_{\alpha}^0$ and $j_{\beta}^0$, respectively.
On the membrane, the concentrations of protein complexes are $C_{\beta}^{\rm{m,d}}$, $C_{\beta}^{\rm{m,a}}$, $C_{\alpha \beta}^{\rm{m,d}}$, $C_{\alpha \beta}^{\rm{m, a}}$---all bound to E-cadherin proteins---which can be either detached or attached to an adjacent cell via E-cadherin-E-cadherin homophilic binding as indicated by the superscripts d and a. Cytoplasmic concentrations at the membrane, denoted
$C_{\beta}^{\rm{b}}$, $C_{\alpha}^{\rm{b}}$ and $C_{\alpha \beta}^{\rm{b}}$, correspond to the respective concentrations introduced in Eq.~\ref{bulkdiffusion} at this particular location. $\beta$-catenin as well as $\alpha$-catenin-$\beta$-catenin complexes---both bound to E-cadherin proteins---can go to the plasma membrane of the cell where they are then in the detached state. We denote by $k^{\rm on}_{\beta}$ and $k^{\rm off}_{\beta}$ ($k^{\rm on}_{\alpha\beta}$ and $k^{\rm off}_{\alpha\beta}$) the rates at which the protein complex $\beta$-catenin-E-cadherin ($\alpha$-catenin-$\beta$-catenin-E-cadherin) goes to the plasma membrane of the cell, and note that only complexes in the detached state can move from the membrane to the cytoplasm. The two fluxes, $j_{\beta}$ and $j_{\alpha \beta}$, of $\beta$-catenin and $\alpha$-catenin-$\beta$-catenin complexes from the cytoplasm to the plasma membrane of the cell then read
\begin{subequations}\label{fluxes1}
\begin{eqnarray}
j_{\beta} &=& k_{\beta}^{\rm on} C_{\beta}^{\rm b} - k_{\beta}^{\rm off} C_{\beta}^{\rm m,d},\\
j_{\alpha \beta} &=& k_{\alpha \beta}^{\rm on} C_{\alpha \beta}^{\rm b} - k_{\alpha \beta}^{\rm off} C_{\alpha \beta}^{\rm m,d}.
\end{eqnarray}
\end{subequations}
Monomeric $\alpha$-catenins can only go to the plasma membrane by forming $\alpha$-catenin-$\beta$-catenin complexes via a reaction with $\beta$-catenin complexes that are already located on the membrane (either in the attached or detached state). The rates of these reactions are denoted respectively by $k_{\alpha \beta}^{\rm m,a}$ and $k_{\alpha \beta}^{\rm m,d}$. Once formed, these complexes do not release pure $\alpha$-catenins anymore. $\alpha$-catenin dimeric complexes on the other hand cannot go to the membrane since they neither attach directly to E-cadherins, nor can they bind $\beta$-catenin-E-cadherin complexes because for that they need to be in their monomeric form. Their flux therefore vanishes. We finally get
\begin{subequations}\label{fluxes2}
\begin{eqnarray}
j_{\alpha} &=& k_{\alpha \beta}^{\rm m,d} C_{\beta}^{\rm m,d} C_{\alpha}^{\rm b} + k_{\alpha \beta}^{\rm m,a} C_{\beta}^{\rm m,a} C_{\alpha}^{\rm b},\\
j_{\alpha\alpha} &=& 0.
\end{eqnarray}
\end{subequations}

\begin{figure}[h]
\includegraphics[width=0.9\linewidth]{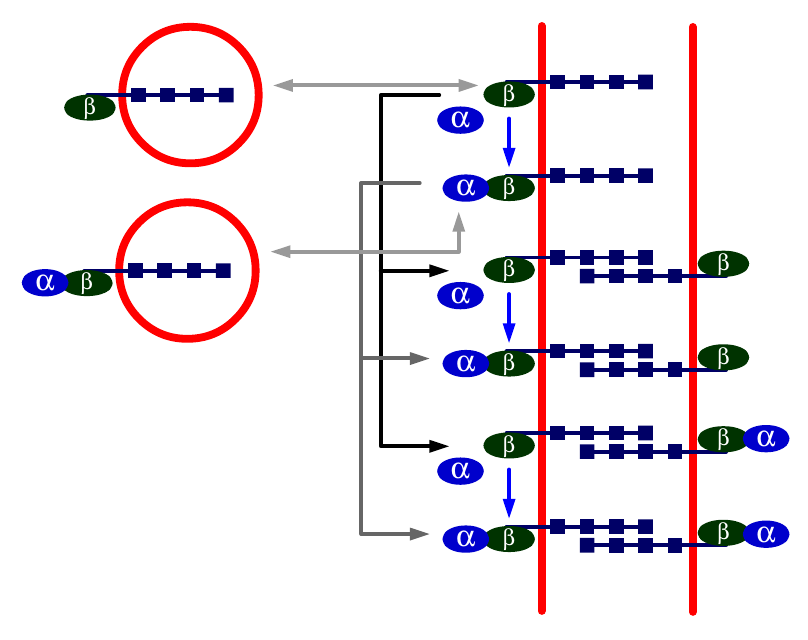}
\caption {\label{fig_membraneReactions}
Schematic illustration of the reactions occurring at the cell membrane and leading to Eq.~\ref{balanceEq}. E-cadherin vesicles, either bound to $\beta$-catenins alone or to $\alpha$-catenin-$\beta$-catenin complexes, can merge with the membrane or be endocytosed. The two associated light gray arrows correspond to the four on and off rates in Eq.~\ref{balanceEq}. In the presence of cell-cell contact, the E-cadherins on the membrane can bind to E-cadherins on the membrane of the adjacent cell, which is represented by the black arrows for $\beta$-catenin-associated complexes, and by the dark gray arrows for $\alpha$-catenin-$\beta$-catenin-associated complexes. These correspond to all the reactions which have rates with an EE superscript in Eq.~\ref{balanceEq}. In addition, $\alpha$-catenin monomers can bind to E-cadherin-$\beta$-catenin complexes present on the membrane, whether they are bound or not to E-cadherins from the adjacent cell. This is represented by the vertical arrows and corresponds to the terms n Eq.~\ref{balanceEq} whose rates have an m,d or m,a superscript. Finally, all protein complexes located on the membrane can be degraded, which is taken into account by the rates labeled with the letter $r$ in Eq.~\ref{balanceEq} (not represented). We assume a symmetric configuration of the adjacent cell.
}
\end{figure}

To solve our system of equations, we must combine these boundary conditions with the cytosolic protein diffusion equations (Eqs.~\ref{bulkdiffusion}a--\ref{bulkdiffusion}d), which we do thanks to the definition of diffusive fluxes ($j_{A} = - D_A \nabla C_A$). To do so, we need to eliminate the membrane protein concentrations from our system of equations. This is done by writing the balance of protein complexes located on the plasma membrane of the cell: In addition to the reaction rates introduced above, we introduce the rate $k^{\rm EE}_{A-B}$ for a given complex $A$ linked to an E-cadherin molecule to attach to another complex $B$ of the adjacent cell via cross-membrane E-cadherin homophilic binding. Also, all complexes are degraded with their specific rates $r^{\rm m,d}_A$ and $r^{\rm m,a}_A$ on the membrane. For simplicity, we assume a completely symmetric, identical configuration of the neighboring cell, and thus identical protein concentrations on the membrane of the adjacent cell. Taking all of this into account, the protein concentrations on the cell membrane are determined by the following steady-state equations, a schematic representation of which is presented in Fig.~\ref{fig_membraneReactions}:
\begin{subequations}\label{balanceEq}
\begin{eqnarray}
&&k_{\beta}^{\rm on} C_{\beta}^{\rm b} - k_{\beta}^{\rm off} C_{\beta}^{\rm m,d} - k_{\alpha  \beta}^{\rm m,d} C_{\beta}^{\rm m,d} C_{\alpha}^{\rm b} - k_{\beta-\beta}^{\rm EE} \left(C_{\beta}^{\rm m,d}\right)^2\nonumber\\
&& \qquad - k_{\alpha \beta -  \beta}^{\rm EE} C_{\beta}^{\rm m,d} C_{\alpha \beta}^{\rm m,d} - r_{\beta}^{\rm m,d} C_{\beta}^{\rm m,d} = 0,\label{balanceEq1}\\
&&k_{\alpha \beta}^{\rm on} C_{\alpha \beta}^{\rm b} - k_{\alpha \beta}^{\rm off} C_{\alpha \beta}^{\rm m,d} + k_{\alpha  \beta}^{\rm m,d} C_{\beta}^{\rm m,d} C_{\alpha}^{\rm b} - k_{\alpha \beta- \alpha \beta}^{\rm EE} \left(C_{\alpha \beta}^{\rm m,d}\right)^2\nonumber\\
&& \qquad - k_{\alpha \beta - \beta}^{\rm EE} C_{\beta}^{\rm m,d} C_{\alpha \beta}^{\rm m,d} - r_{\alpha \beta}^{\rm m,d} C_{\alpha \beta}^{\rm m,d} = 0,\label{balanceEq2}\\
&&- k_{\alpha \beta}^{\rm m,a} C_{\beta}^{\rm m,a} C_{\alpha}^{\rm b} + k_{\beta-\beta}^{\rm EE} \left(C_{\beta}^{\rm m,d}\right)^2 + k_{\alpha \beta- \beta}^{\rm EE} C_{\beta}^{\rm m,d} C_{\alpha \beta}^{\rm m,d}\nonumber\\
&&  \qquad - r_{\beta}^{\rm m,a} C_{\beta}^{\rm m,a} = 0,\label{balanceEq3}\\
&&k_{\alpha \beta}^{\rm m,a} C_{\beta}^{\rm m,a} C_{\alpha}^{\rm b} + k_{\alpha \beta -\alpha \beta}^{\rm EE} \left(C_{\alpha \beta}^{\rm m,d}\right)^2 + k_{\alpha \beta- \beta}^{\rm EE} C_{\beta}^{\rm m,d} C_{\alpha \beta}^{\rm m,d}\nonumber\\
&& \qquad - r_{\alpha \beta}^{\rm m,a} C_{\alpha \beta}^{\rm m,a} = 0\label{balanceEq4}.
\end{eqnarray} 
\end{subequations}
Here, the two first equations describe the balance of $\beta$-catenin and $\alpha$-catenin-$\beta$-catenin complexes, respectively, on the plasma membrane of the cell that are detached from the neighboring cell, and the two following equations do the same for the attached protein complexes. For example, in the first equation, $\beta$-catenin-E-cadherin complexes in the detached state can---in the order of the terms present in the equation---be replenished via attachment of $\beta$-catenin-E-cadherin complexes from the cytoplasm, disappear via endocytosis of $\beta$-catenin-E-cadherin complexes, form $\alpha$-catenin-$\beta$-catenin-E-cadherin complexes, attach with other $\beta$-catenin-E-cadherin complexes from the neighboring cell via cross-membrane E-cadherin-E-cadherin homophilic binding, or with $\alpha$-catenin-$\beta$-catenin-E-cadherin complexes from the neighboring cell, and, finally, disappear via degradation. The terms in the second equation are of similar origin. The third and fourth equations for the attached states resemble the previous ones, except that there is no exchange of proteins directly with the cytoplasm in this case.

\subsection*{Steady-State Concentration Profiles}

To solve the system of equations given by Eqs.~\ref{bulkdiffusion}--\ref{balanceEq}, we now separate the two cases of presence and absence of contact with a neighboring cell, for which we can separately eliminate the membrane concentrations from the boundary conditions given by Eqs.~\ref{fluxes1} and \ref{fluxes2} thanks to Eq.~\ref{balanceEq}. In the absence of cell-cell contact, the different rates $k^{\rm EE}_{A-B}$ vanish and all the proteins on the membrane are in the detached state. In this case, Eqs.~\ref{balanceEq3} and \ref{balanceEq4} become irrelevant, and Eqs.~\ref{balanceEq1} and \ref{balanceEq2} become
\begin{subequations}
\begin{eqnarray}
k_{\beta}^{\rm on} C_{\beta}^{\rm b} - k_{\beta}^{\rm off} C_{\beta}^{\rm m,d} - k_{\alpha  \beta}^{\rm m,d} C_{\beta}^{\rm m,d} C_{\alpha}^{\rm b} - r_{\beta}^{\rm m,d} C_{\beta}^{\rm m,d} &=& 0,\label{balanceEq1Simpl}\nonumber\\
&&
\end{eqnarray} 
and
\begin{eqnarray}
k_{\alpha \beta}^{\rm on} C_{\alpha \beta}^{\rm b} - k_{\alpha \beta}^{\rm off} C_{\alpha \beta}^{\rm m,d} + k_{\alpha  \beta}^{\rm m,d} C_{\beta}^{\rm m,d} C_{\alpha}^{\rm b} - r_{\alpha \beta}^{\rm m,d} C_{\alpha \beta}^{\rm m,d} &=& 0.\label{balanceEq2Simpl}\nonumber\\
&&
\end{eqnarray} 
\end{subequations}
The first equation allows to solve for $C_{\beta}^{\rm m,d}$ and then express $j_{\alpha}$ and $j_{\beta}$ as a function of the cytosolic concentrations of $\alpha$-catenin and $\beta$-catenin complexes only.
Thus, we get a closed set of equations for these two quantities, in which we find the cytosolic equations (Eqs.~\ref{Eq1} and \ref{Eq2}), as well as the expressions for the fluxes at the boundaries of the system, namely the imposed fluxes $j_{\alpha}^0$ and $j_{\beta}^0$ at the Golgi apparatus of the cell and the two following fluxes at the cell membrane,
\begin{subequations}\label{B1}
\begin{eqnarray}
j_{\beta} &=& \frac{r_{\beta}^{\rm m,d} + k_{\alpha \beta}^{\rm m,d}C_{\alpha}^{\rm b}}{r_{\beta}^{\rm m,d} + k_{\beta}^{\rm off} + k_{\alpha \beta}^{\rm m,d} C_{\alpha}^{\rm b}}\,k_{\beta}^{\rm on}C_{\beta}^{\rm b}, \label{BC1}\\
j_{\alpha} &=& \frac{k_{\alpha \beta}^{\rm m,d} C_{\alpha}^{\rm b}}{r_{\beta}^{\rm m,d} + k_{\beta}^{\rm off} + k_{\alpha \beta}^{\rm m,d} C_{\alpha}^{\rm b}}\,k_{\beta}^{\rm on}C_{\beta}^{\rm b}. \label{BC2}
\end{eqnarray}
\end{subequations}
This system can be solved independently and then used to solve for the concentrations of $\alpha$-catenin-$\beta$-catenin and $\alpha$-catenin dimeric complexes in a second step, using the remaining equations.

In the presence of cell-cell contact, we assume for simplicity $k^{\rm EE} \rightarrow \infty$, meaning that all protein complexes on the membrane instantaneously bind to the neighboring cell. Therefore, the concentrations of unbound proteins on the membrane $C_{\beta}^{\rm m,d}$ and $C_{\alpha \beta}^{\rm m,d}$ vanish. Similarly to the previous case, the dynamics for $\alpha$-catenin and $\beta$-catenin complexes decouples from the rest of the system, and the fluxes can be obtained from Eqs.~\ref{fluxes1} and \ref{fluxes2} after we have solved for $C_{\beta}^{\rm m,a}$ using Eqs.~\ref{balanceEq1} and \ref{balanceEq3}:
\begin{subequations}\label{B2}
\begin{eqnarray}
j_{\beta} &=& k_{\beta}^{\rm on}C_{\beta}^{\rm b}, \label{BC4}\\
j_{\alpha} &=& \frac{k_{\alpha \beta}^{\rm m,a} C_{\alpha}^{\rm b}}{r_{\beta}^{\rm m,a} + k_{\alpha \beta}^{\rm m,a} C_{\alpha}^{\rm b}}\,k_{\beta}^{\rm on}C_{\beta}^{\rm b}. \label{BC5}
\end{eqnarray}
\end{subequations}

\begin{figure}[h]
\scalebox{0.65}{
\includegraphics[width=0.9\linewidth]{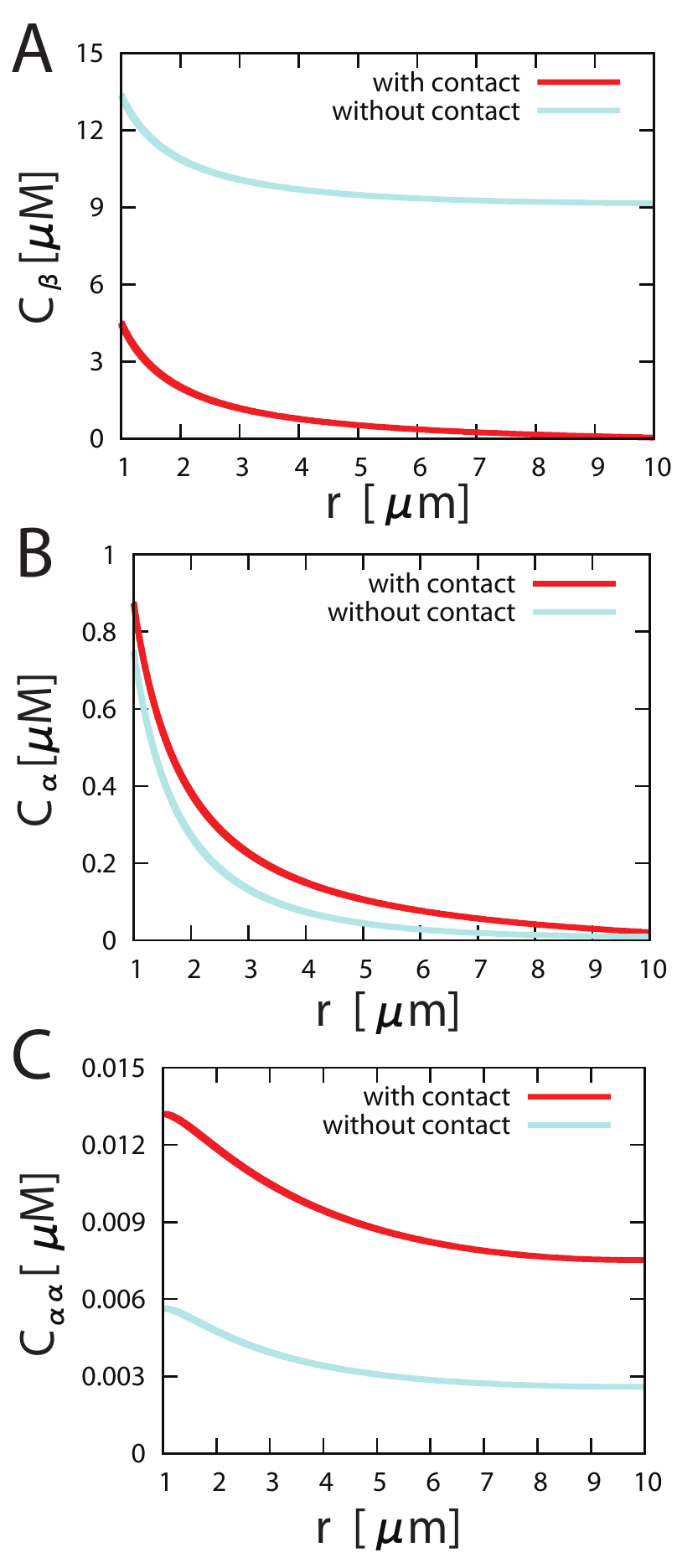}
}
\caption {\label{fig_conc}
Cytosolic concentration profiles of $\beta$-catenin (A), $\alpha$-catenin (B) and $\alpha$-catenin dimers (C) resulting from the reaction-diffusion system described by Eqs.~\ref{bulkdiffusion}a--\ref{bulkdiffusion}d, with the boundary conditions without cell-cell contact (Eq.~\ref{B1}) or with cell-cell contact (Eq.~\ref{B2}), as functions of the distance from the center of a cell with spherical symmetry. The different parameters are as follows. The Golgi apparatus of the cell is located at $r_0=1$~$\mu$m and the total radius of the cell is $R=10$~$\mu$m. The $\beta$-catenin influx is $j^0_{\beta}=5.0~\mu$m~$\mu$M~s$^{-1}$. The diffusion constants of $\alpha$- and $\beta$-catenin are equal to $1$~$\mu$m$^2$s$^{-1}$, and the one for $\alpha$-catenin dimers is $0.5$~$\mu$m$^2$s$^{-1}$ \citep{konopka2006cac}. In these plots, the reaction rate of $\alpha$-catenin with $\beta$-catenin is $0.01$~$\mu$M$^{-1}$s$^{-1}$, and the one of $\alpha$-catenin with itself is $0.005$~$\mu$M$^{-1}$s$^{-1}$ \citep{schlosshauer2004rpp}. The protein degradation rates of $\alpha$- and $\beta$-catenin are equal to $10^{-3}$~s$^{-1}$ and the consumption and degradation rates of $\alpha$-catenin dimers are $0.5$~$10^{-3}$~s$^{-1}$. Finally, the membrane binding and unbinding rates $k^{\rm on}_{\beta}$ and $k^{\rm off}_{\beta}$ of the $\beta$-catenin protein complex are equal to 1~$\mu$m s$^{-1}$ and 1~s$^{-1}$, respectively.
}
\end{figure}

The system of equations derived above can now be solved independently in the two configurations of the cell numerically, namely in the presence or absence of contact with a neighboring cell. It consists of Eqs.~\ref{Eq1} and \ref{Eq2}, together with the boundary conditions (Eq.~\ref{B1} or \ref{B2}) at the plasma membrane and a constant protein influx given by $j_{\alpha}^0$ and $j_{\beta}^0$ at the Golgi apparatus of the cell. We solve this system for the case of a spherical cell of radius $R$, whose Golgi apparatus is modeled as a sphere of radius $r_0$. In Fig.~\ref{fig_conc}, we illustrate the difference in the concentrations of $\alpha$-catenin, $\beta$-catenin and $\alpha$-catenin dimers with and without cell-cell contact.
We see in Fig.~\ref{fig_conc}~C that the overall concentration of $\alpha$-catenin dimers presents a significant increase in the case of cell-cell contact as compared to the case without contact, which provides an efficient switch between the two phenotypic states of the cell. Note that in both cases, there is a drop in the concentration of $\alpha$-dimers away from the nucleus. If the diffusion constant is small enough (i.e. if $D_{\alpha\alpha}/[R^2(r_{\alpha\alpha}+\tilde{r}_{\alpha\alpha})] \ll 1$), this concentration drop is significant and could be relevant for the spatial organization of polymerized actin within the cell. We comment further on this aspect in the discussion section.

\subsection*{Scaling Analysis}

Let us now perform a scaling analysis of the total number of $\alpha$-catenin dimers in the system as given by our model, comparing the two cases with and without cell-cell contact. In asymptotic limits in which the involved length scales separate, it is possible to solve our system of equations analytically. We thereby obtain a better physical understanding of the contact inhibition mechanism proposed in this paper. We also derive a simple expression for the change in the total amount of $\alpha$-catenin dimers $N_{\alpha \alpha}$ in the cell between the contact and no-contact states, which dictates the amplitude of the switch. This final expression is given by Eq.~\ref{ratio}, and one may want to skip to this equation and its associated comments directly. Later, this derivation also helps us to see in which biological conditions our mechanism can function, as well as to investigate the various possibilities that can lead to its breakdown. This is done in the next section.

We first identify the different characteristic lengths over which the concentrations of the different proteins vary as each of the reactions is considered separately. For a given protein, the shortest of the characteristic lengths of the different reactions it enters determines its dominant reaction pathway. The characteristic length for the change in monomeric $\alpha$-catenin concentration due to $\alpha$-catenin dimer formation ($\alpha$-catenin-to-$\beta$-catenin binding) is given by $l^{\alpha}_{\alpha \alpha} = \sqrt{D_{\alpha}/(k_{\alpha \alpha} C_{\alpha})}$ ($l^{\alpha}_{\alpha \beta} = \sqrt{D_{\alpha}/(k_{\alpha\beta}C_{\beta})}$). In a similar way, the change in $\beta$-catenin concentration due to
$\alpha$-catenin-to-$\beta$-catenin binding is given by $l^{\beta}_{\alpha \beta} = \sqrt{D_{\beta}/(k_{\alpha\beta}C_{\alpha})}$. Finally, the characteristic length due to monomeric $\alpha$-catenin degradation ($\beta$-catenin degradation) is $l_{\alpha} = \sqrt{D_{\alpha}/r_{\alpha}}$ ($l_{\beta} = \sqrt{D_{\beta}/r_{\beta}}$).

We first look at the case where $l^{\alpha}_{\alpha \beta}$ is the shortest of the length scales given above. As we shall see below, such a condition is realized as soon as the production of $\beta$-catenin at the Golgi apparatus of the cell is large-enough, such that reactions with monomeric $\beta$-catenin proteins (bound to E-cadherins) are fast. In this case, $\alpha$-catenin-to-$\beta$-catenin binding is dominant over $\alpha$-catenin dimerization in the absence of cell-cell contact, such that our mechanism can be efficient. (Other limits are studied below.) The concentration of $\alpha$-catenin at the cell membrane is always very low and the change in the steady state concentration of $\alpha$-catenin with and without cell-cell contact results from a redistribution of $\beta$-catenin inside the cell. 

Within this limit, we can assume a quasi-constant concentration of $\beta$ catenin in a region of length $l^{\alpha}_{\alpha \beta}$ around its source, which allows us to find analytical expressions for the reaction-diffusion system in a one-dimensional geometry with coordinate $x$, the protein source being at $x=r_0$ and the cell membrane at $x=R$. The solution for the $\alpha$-catenin concentration is given by:
\begin{eqnarray} \label{alphasol}
C_{\alpha} \simeq C_{\alpha}^0 \exp\left(-(x-r_0)/l^{\alpha}_{\alpha \beta}\right),
\end{eqnarray}
with $C_{\alpha}^0 = l^{\alpha}_{\alpha \beta}\,j^{0}_{\alpha}/D_{\alpha} = j^{0}_{\alpha}/\sqrt{k_{\alpha \beta} C_{\beta}^0 D_{\alpha}}$, 
and where $C_{\alpha}^0$ and $C_{\beta}^0$ are the concentrations of $\alpha$-catenin and $\beta$-catenin complexes, respectively, at $x=r_0$. The solution for the $\beta$-catenin concentration is given by
\begin{eqnarray} \label{betasol}
C_{\beta} \simeq C_{\beta}^0 \cosh{\left( (x-r_0)/l_{\beta} \right)} - \frac{j^{0}_{\beta}-j^{0}_{\alpha}}{\sqrt{D_{\beta} r_{\beta}}} \sinh{\left( (x-r_0)/l_{\beta} \right)}.
\end{eqnarray}
From the boundary conditions, we can determine the expression for $C_{\beta}^0$: 
\begin{eqnarray} \label{beta0}
C_{\beta}^0 \simeq f\,\frac{j^{0}_{\beta}-j^{0}_{\alpha}}{\sqrt{D_{\beta} r_{\beta}}},
\end{eqnarray}
where 
\begin{eqnarray}\label{f&g}
f = \frac{ 1 + g\,\tanh{\left( R/l_{\beta} \right)} }{\tanh{\left( R/l_{\beta} \right)} + g}
\quad
\textrm{and}
\quad
g=\frac{r_{\beta}^{\rm m,d}}{k^{\rm off}_{\beta} + r_{\beta}^{\rm m,d}}\,\frac{k^{\rm on}_{\beta}}{\sqrt{D_{\beta}r_{\beta}}}.
\end{eqnarray}

It is now possible to translate our initial assumptions on the different characteristic reaction lengths into conditions directly on the concentration $C_{\beta}^0$ of $\beta$-catenin at $x=r_0$. For the three characteristic lengths $l_i=l_\alpha, l_\beta$ and $R$, the condition $l^{\alpha}_{\alpha \beta}\ll l_i$ reads
\begin{eqnarray}
C_{\beta}^0\gg \frac{D_\alpha}{k_{\alpha\beta}}\,\frac{1}{l_i^2},
\end{eqnarray}
while the conditions $l^{\alpha}_{\alpha \beta}\ll l^{\beta}_{\alpha \beta}$ and $l^{\alpha}_{\alpha \beta}\ll l^{\alpha}_{\alpha \alpha}$, respectively, read
\begin{eqnarray}
C_{\beta}^0\gg \left[\frac{D_\alpha}{k_{\alpha\beta}}\left(\frac{j^0_{\alpha}}{D_\beta}\right)^2\right]^{1/3},
\end{eqnarray}
and
\begin{eqnarray}
C_{\beta}^0\gg \frac{\left(2k_{\alpha\alpha}j^0_{\alpha}\right)^{2/3}}{k_{\alpha\beta}\left( D_{\alpha}\right)^{1/3}}.
\end{eqnarray}
Finally, there is an additional condition stating that the concentration of $\beta$-catenin is quasi-constant in a region of length $l^{\alpha}_{\alpha \beta}$ around the protein source:
\begin{eqnarray}
C_{\beta}^0\gg \frac{D_\alpha r_\beta}{k_{\alpha\beta}D_\beta}\,\frac{1}{f^2}.
\end{eqnarray}
Since $C_{\beta}^0 \propto j^0_{\beta}-j^0_{\alpha}$, all of these conditions are satisfied for a sufficiently large influx $j^0_{\beta}$ of $\beta$-catenins at the Golgi apparatus of the cell. 

The amplitude of the switch is given by the change in the total amount of $\alpha$-catenin dimers $N_{\alpha \alpha}$ in the cell between the contact and no-contact states. The concentration in $\alpha$-catenin dimers is simply given by
\begin{eqnarray}
C_{\alpha \alpha} = \frac{k_{\alpha \alpha}}{r_{\alpha \alpha}+\tilde{r}_{\alpha \alpha}}\,C_{\alpha}^2,
\end{eqnarray}
if $l^{\alpha}_{\alpha \beta}$ is much smaller than the two other lengths scales given by Eq.~\ref{Eq3}, namely $\tilde{l}_{\alpha \alpha}^{\alpha}=\sqrt{D_{\alpha \alpha}C_{\alpha \alpha}/(k_{\alpha \alpha}C_{\alpha}^2)}$ and $l_{\alpha \alpha}=\sqrt{D_{\alpha \alpha}/(r_{\alpha \alpha}+\tilde{r}_{\alpha \alpha})}$.
Integrating the $\alpha$-catenin dimer concentration over the size of the whole cell under this hypothesis, we obtain
\begin{eqnarray} \label{Ndimers}
N_{\alpha \alpha} \simeq \frac{k_{\alpha \alpha} (j^0_{\alpha})^2 } {2 r_{\alpha \alpha} (k_{\alpha \beta})^{3/2} (D_{\alpha})^{1/2}}\,(C_{\beta}^0)^{-3/2}
\end{eqnarray}
as a formal expression.
The consistency check for this expression gives the following condition on $C_{\beta}^0$:
\begin{eqnarray}
C_{\beta}^0\gg \frac{D_\alpha k_{\alpha\beta}}{D_{\alpha\alpha}(r_{\alpha \alpha}+\tilde{r}_{\alpha \alpha})},
\end{eqnarray}
which again is satisfied for a sufficiently large influx $j^0_{\beta}$ of $\beta$-catenins at the Golgi apparatus of the cell. The results in the presence and absence of cell-cell contact can be obtained by switching the rates for detached membrane proteins with those for attached ones. In particular, the off-rate of $\beta$-catenin from the membrane, $k^{\rm off}_\beta$, must be set to $0$ in the case where there is contact with a neighboring cell. If all the other rates stay the same, we obtain a simple expression for the ratio of the total amounts of $\alpha$-catenin dimers with and without cell-cell contact:
\begin{eqnarray} \label{ratio}
\frac{N_{\alpha \alpha}^{\rm con}}{N_{\alpha \alpha}^{\rm no con}} \simeq \left( 1 + \frac{k^{\rm off, nocon}_{\beta}}{r_{\beta}^{\rm m,d}}\right)^{3/2},
\end{eqnarray}
which comes from the dependence of $N_{\alpha \alpha}$ on $C_{\beta}^0$, and where $k^{\rm off, nocon}_{\beta}$ is the off-rate of $\beta$-catenin from the membrane when their is no cell-cell contact. Hence, for a protein degradation rate $r_{\beta}^{\rm m,d}$ much smaller than the off rate $k^{\rm off}_{\beta}$, we expect a significant switch in the total amount of $\alpha$-catenin dimers produced and, thus, a functional contact inhibition mechanism. 

We now look at two cases where $l^{\alpha}_{\alpha \beta}$ may not necessarily be the smallest characteristic length in the system. First, it is possible that, when there is cell-cell contact, $l^{\alpha}_{\alpha \alpha}$ becomes the shortest characteristic length instead of $l^{\alpha}_{\alpha \beta}$ in the absence of contact. Indeed, when there is cell-cell contact, $\beta$-catenin proteins could be sufficiently depleted from the cell for most of the $\alpha$-catenins to form homo-dimeric complexes before reacting with $\beta$-catenins. In this case, the contact inhibition switch remains intact, and the previous ratio still scales as stated in Eq.~\ref{ratio}.

\begin{figure}[h]
\includegraphics[width=\linewidth]{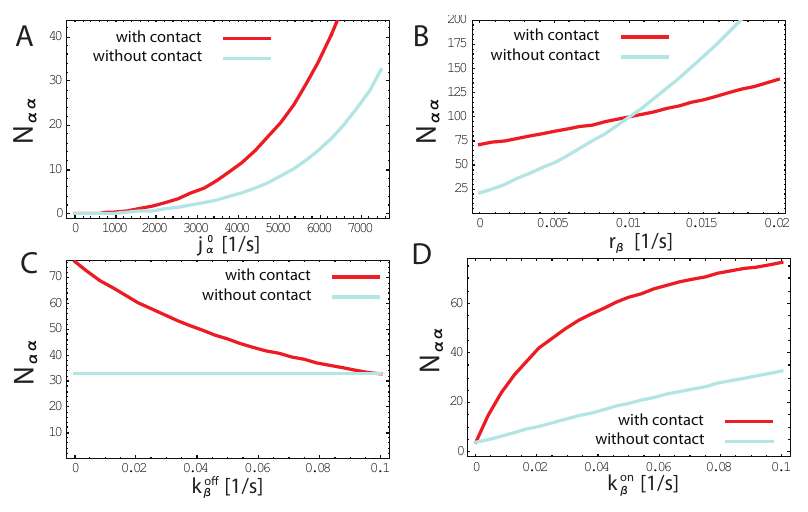}
\caption {\label{fig_0D}
Numerical solutions for the total number of $\alpha$-catenin dimers in the cell as a function of different parameters when the cadherin-catenin system is treated as a zero-dimensional reaction system. In this limit, the contact inhibition mechanism is based on a larger degradation rate on the membrane ($r_{\beta}^{\rm m,d}=10^{-2} $~s$^{-1}$) than in the cytosol ($r_{\beta}=10^{-3} $~s$^{-1}$). The $\alpha$-catenin degradation rate $r_{\alpha}$ is also assumed to be small ($10^{-3}$~s$^{-1}$). The total production rates of $\alpha$-catenin and $\beta$-catenin are respectively equal to 7.5~$10^3$~s$^{-1}$ and 14~$10^3$~s$^{-1}$. Protein-protein reaction rates are $10^{-3}$ s$^{-1}$, and on- and off-rates of the $\beta$-catenin protein complex to the cell membrane are $k^{\rm on}_{\beta}=k^{\rm off}_{\beta}=0.1$~s$^{-1}$. Note that for $r_{\beta}>r_{\beta}^{\rm m,d}$, the switch is reversed.
}
\end{figure}

Another limit corresponds to the case where the cell radius, $R$, is the shortest length scale in the system. In this case, the system of reaction diffusion equations---together with the corresponding boundary conditions---can be treated as a system without spatial extension. A substantial change in the $\beta$-catenin concentration between the two states of the cell can be achieved in this limit if the degradation rate of $\beta$-catenin in the cytosol $r_{\beta}$ is much smaller than its degradation rate on the membrane $r_{\beta}^{\rm m,d}$. We present the numerical solutions that correspond to this limit in Fig.~\ref{fig_0D}.
In particular, Fig.~\ref{fig_0D}~B shows that the switch is controlled by the ratio $r_{\beta}/r_{\beta}^{\rm m,d}$. As an alternative, the reaction rate $k_{\alpha \beta}^{\rm m, a}$ of membrane-bound $\beta$-catenin with $\alpha$-catenin could be smaller than the corresponding reaction rate in the cytosol $k_{\alpha \beta}$. This would also yield a functioning contact inhibition switch.

\subsection*{Breakdown of Contact Inhibition}

\begin{figure}[h]
\includegraphics[width=\linewidth]{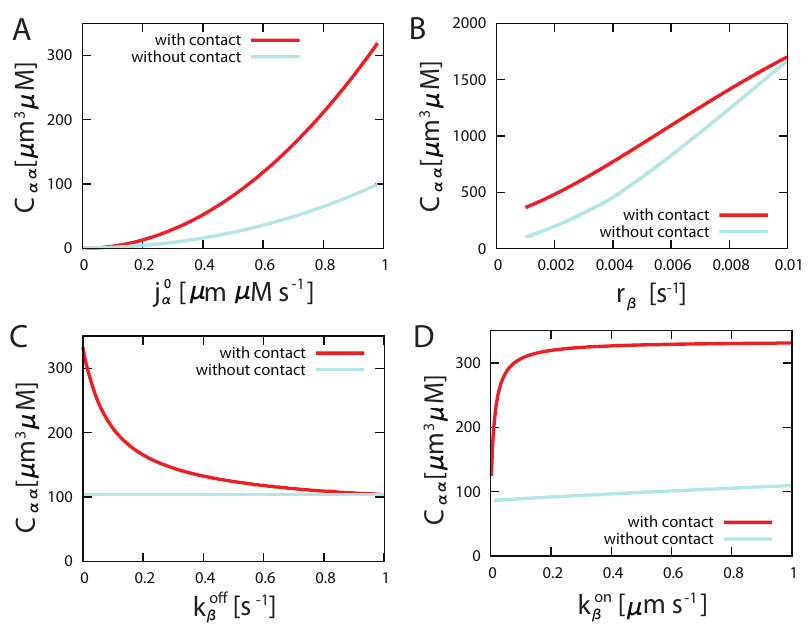}
\caption {\label{fig_break}
Integrated $\alpha$-catenin dimer concentration over the cell volume from the numerical solution of the whole model. This quantity is plotted as a function of the $\alpha$-catenin influx $j^0_{\alpha}$ (A), of the $\beta$-catenin degradation rate in the cytoplasm $r_{\beta}$ (B) and of the unbinding and binding rates of the $\beta$-catenin protein complex off and on the membrane $k^{\rm off}_{\beta}$ (C) and $k^{\rm on}_{\beta}$ (D). The same constants are used as in Fig.~\ref{fig_conc}. In B, $r^{\rm{m,d}}_{\beta}=r_\beta$ is assumed. Contact inhibition occurs when the concentration of $\alpha$-catenin dimers $C_{\alpha\alpha}$ is large. The contact-inhibition breaks down for low production $j^0_{\alpha}$ of $\alpha$-catenin (A) and for an increased degradation of $\beta$-catenin (B). It also breaks down for an increased membrane off-rate (C), which corresponds to mutations of E-cadherins leading to less efficient formation of cell-cell E-cadherin bonds. This results in an insufficient trapping of $\beta$-catenins on the membrane in the presence of cell-cell contact. Finally, contact inhibition breaks down for small values of $k^{\rm on}_{\beta}$ (D), which could correspond to a less efficient binding of $\beta$-catenin to the membrane due to a decreased expression of E-cadherins.
}
\end{figure}

Let us now investigate the various possibilities that, according to our model, can lead to a breakdown of the contact inhibition pathway. To investigate what affects the production of $\alpha$-catenin dimers, let us look at the total amount of $\alpha$-catenin dimers in the cell calculated from a numerical solution of our whole system of equations (Eqs.~\ref{bulkdiffusion}a--\ref{bulkdiffusion}d) together with the boundary conditions described by Eqs.~\ref{B1} and \ref{B2}, for the contact-free and contact-inhibited states (see Fig.~\ref{fig_break}).
We first show the dependence on $j_{\alpha}^0$ (Fig.~\ref{fig_break}~A), which is proportional to the total production of $\alpha$-catenin in the cell Golgi apparatus. As discussed above, many cancerous cells show mutations impairing the function or production of $\alpha$-catenin proteins. From Eq.~\ref{Ndimers}, we see that the total amount of $\alpha$-catenin dimers scales like $(j^0_{\alpha})^2$, which is consistent with these experimental observations. For low values of $j_{\alpha}^0$, the difference between the contact-free and contact-inhibited state disappears, as it has been experimentally observed \cite{vasioukhin2001had}. Next, we consider the effect of a knockout of the APC protein, which is known to label the $\beta$-catenin in the cytosol for degradation. From Eqs.~\ref{beta0}, \ref{f&g}, and \ref{Ndimers}, we see that for a fast $\beta$-catenin degradation, the total amount of $\alpha$-catenin dimers scales like $(r_{\beta})^{3/4}$. Hence, contact inhibition becomes less effective for a lower cytosolic $\beta$-catenin degradation rate, as it has been observed in experiments~\cite{korinek1997cta}. As can be seen in Fig.~\ref{fig_break}~B, a low degradation rate of $\beta$-catenin---which corresponds to a knockout of APC---leads to a total concentration of $\alpha$-catenin dimers in the cell that is low, even when there is cell-cell contact. Other defects frequently encountered for cancerous cells are downregulation or mutations of E-cadherins \cite{hirohashi1998icm}. In our picture, a malfunction of E-cadherins due to mutations corresponds to a less effective binding of E-cadherins to neighboring cells, and thereby a less efficient trapping of E-cadherins and $\beta$-catenins to the plasma membrane. Fig.~\ref{fig_break}~C shows that an increased off-rate $k^{\rm off}_\beta$ in both states again leads to a failure of contact inhibition, since the difference between the contact and no-contact states disappears for large values of $k^{\rm off}_\beta$ in the contact state. Finally, a lower expression of E-cadherin results in a less effective binding of $\beta$-catenin to the plasma membrane, which can be modeled by a decreased on-rate $k^{\rm on}_\beta$, as shown in Fig.~\ref{fig_break}~D. Similar results are obtained when $R$ is the shortest length scale in the system. In this case, the solution of the system of equations is plotted in Fig.~\ref{fig_0D} as a function of the model parameters that simulate a breakdown of the contact inhibition mechanism. The breakdown of the switch is similar to the one discussed above.

\section*{Discussion}

In this article, we have proposed a reaction-diffusion model of the cadherin-catenin system in which the concentration of $\alpha$-catenin dimers increases in a confluent cell as compared to a cell without contact. We propose that this switch is due to a competition between mutually exclusive $\alpha$-catenin dimerization and $\alpha$-catenin-to-$\beta$-catenin binding in the cytosol of the cell. In the presence of cell-cell contact, intercellular E-cadherin bonds prevent endocytosis of E-cadherin complexes. This leads to a redistribution of unbound $\beta$-catenins to the cell membrane and thereby a significant increase in the amount of $\alpha$-catenin dimerization. Hence, the cell shifts between an active state combining high cellular expansion pressure with high cellular motility, to a quiescent state, where actin branching is inhibited. 

From our analysis, we expect the contact inhibition switch to function efficiently if $\beta$-catenin is sufficiently abundant in the cell to effectively compete with $\alpha$-catenin dimerization. Therefore, there are three distinct possibilities that can lead to a functioning contact inhibition mechanism. First, the protein reaction rates could be sufficiently fast compared with protein diffusion to effectively separate two distinct pools of $\beta$-catenin proteins, respectively cytosolic and membrane-bound $\beta$-catenins (both linked to E-cadherin proteins). While $\beta$-catenins in the cytosol compete with $\alpha$-catenin dimerization, $\beta$-catenins on the membrane cannot react with $\alpha$-catenins anymore, because all $\alpha$-catenins either bind $\beta$-catenins or form dimers before they get a chance to arrive at the membrane. In that case, the contact inhibition switch comes from a redistribution of $\beta$-catenin-E-cadherin complexes from the cytosol toward the cell membrane as a response to contact with a neighboring cell, effectively letting $\alpha$-catenin dimers form in the cytosol before reaching the cell membrane, where a high concentration of $\beta$-catenin proteins is found. Second, the contact-inhibition switch could arise from depletion of $\beta$-catenins from the entire cell in the state with contact as compared to the state without contact. This is case for example if the degradation rate of $\beta$-catenin is much larger on the membrane than it is in the cytosol, which is possible since $\beta$-catenin degradation takes place via two distinct pathways in the cytosol and on the membrane of the cell. In this case, the mechanism works even for very slow reaction rates and fast protein diffusion, and the spatial structure of the cell can then be ignored or be treated as a zero-dimensional system. This case is illustrated in Fig.~\ref{fig_0D}. Third, the switch could arise from a reaction of membrane-bound $\beta$-catenin with $\alpha$-catenin much slower than the corresponding reaction rate in the cytosol. Depending on the state of the cell, $\beta$-catenin proteins are indeed located primarily either on the plasma membrane or in the cytosol, allowing $\alpha$-catenin proteins to dimerize or not. This case also does not rely on slow diffusion and can be treated as a zero-dimensional system. 

In addition to providing a mechanism for contact inhibition, the model qualitatively reproduces the effect of several mutations that are known to cause the breakdown of this mechanism and result in tumor-like phenotypes. As we have seen from Figs.~\ref{fig_0D} and \ref{fig_break}, the model agrees with experimental observations in which the expression level or the degradation rate of either E-cadherin, $\beta$-catenin or $\alpha$-catenin are modified. These findings explain why a broad range of mutations leads to similar cancerous phenotypes. In particular, the effects of an increased $\beta$-catenin concentration on contact inhibition are explained by this model without implicating the Wnt-signaling pathway~\citep{stockinger2001crc}.

While this work shows that the cadherin-catenin reaction-diffusion system could play the role of a contact inhibition switch, it is impossible to determine whether it is the most relevant effect without further experimental studies. Experiments that would directly test this pathway are possible. We have already discussed how the cadherin-catenin mechanism reproduces the observed effects of a change in the production and degradation rates of different proteins. However, it might also be possible to only inhibit the interaction of any given pair of these proteins by phosphorylation of specific residues, and thereby directly investigate every step in the proposed mechanism without interfering with other pathways like the Wnt-signaling pathway~\citep{clevers2006wnt, stockinger2001crc}. For example, one experiment of particular interest would be to knockout $\beta$-catenin-to-$\alpha$-catenin binding by phosphorylation without changing the level of expression of these proteins. This would distinguish our model from the picture proposed by Nelson et al.~\citep{drees2005alphacm, yamada2005dcc}: indeed, within the reaction-diffusion model presented here, such a treatment would result in an increased concentration of $\alpha$-catenin dimers, the inhibition of actin branching and thus the contact-inhibited cell state even in the absence of a neighboring cell. In contrast, in the picture proposed by Nelson et al., failure of $\beta$-catenin to bind to $\alpha$-catenin would lead to a disruption of the localization of $\alpha$-catenin to the adhesion sites, and thereby to actin branching and polymerization even at confluence. 

If the model presented here were to be confirmed experimentally, it could potentially ground the idea that different homeostatic growth pressures between neoplastic and healthy tissues are responsible for tumor growth~\cite{basan09}. In particular, one could then test whether different disruptions of the cadherin-catenin pathway that are known to lead to tumorigenesis would affect the homeostatic pressure growths of the tissues under study. Such an observation could potentially give a direct explanation of the observed link between the cadherin-catenin system and neoplastic phenotypes.

\section*{Acknowledgments}

We thank F. Amblard, J. Whitehead, E. Farge and C. Storm for a critical reading of the manuscript and many helpful discussions. T.I. acknowledges financial support from the Fundation for Fundamental Research on Matter of the Netherlands Organisation for Scientific Research (NWO-FOM) within the program on Material Properties of Biological Assemblies (grant FOM-L2601M).

\end{document}